\documentclass[pra,aps,twocolumn]{revtex4} 
\usepackage[dvipdfmx]{graphicx}
\usepackage{amsmath,amssymb,amsfonts}
\usepackage{bm}
\usepackage{subfigure}
\def\be{\begin{equation}}
\def\ee{\end{equation}}
\def\ket{\rangle}
\def\bra{\langle}

\begin{document}

\title{Quantum Error Correction with Uniformly Mixed State Ancillae}
 
\author{Yasushi Kondo,$^{1,2}$ Chiara Bagnasco,$^{1}$ 
and Mikio Nakahara$^{1,2}$}
  
\affiliation{$^1$Research Center for Quantum Computing,
Interdisciplinary Graduate School of Science and Engineering,
Kinki University, Higashi-Osaka, 577-8502, Japan\\
$^2$Department of Physics,
Kinki University, Higashi-Osaka, 577-8502, Japan
}

\begin{abstract}
It is often assumed that the ancilla qubits required for encoding a qubit 
in quantum error correction (QEC) have to be in pure states, 
$|00 \dots 0\rangle$ for example. In this letter, 
we introduce an encoding scheme avoiding fully correlated errors, 
in which the ancillae may be in a uniformly mixed
state. We demonstrate our scheme experimentally by making use of 
a three-qubit NMR quantum computer. Moreover, 
the encoded state has an interesting nature in terms of Quantum Discord, 
or purely quantum correlations between the data-qubit and the ancillae. 
\end{abstract}

\maketitle 

A quantum computer is vulnerable against environmental
noise and it must be protected by one way or another.
Quantum error correction
(QEC) is one of the most successful approaches to this end
\cite{gaitan}. 
Despite this great success, QEC requires 
expensive resources, or ancillae that are
usually assumed to be in pure states \cite{divin, pure_a}. 
However, it is not yet proved that ancillae in uniformly mixed states
are useless. We extend previous works \cite{qecc3} and 
show an encoding scheme robust against fully correlated noise
in which all the ancillae can be in uniformly mixed states.
The encoded state has an interesting nature in terms of 
Quantum Discord \cite{QD}, 
or purely quantum correlations between the data-qubit and the ancillae. 
Our QEC scheme also provides an example of 
Deterministic Quantum Computation with 1-Qubit (DQC-1) 
\cite{dqc1,Zeeya}.

Suppose we have a single qubit in an arbitrary state $\rho_1$, 
which we want to protect from noise. 
We introduce some additional qubits (ancillae) in order to protect the
first qubit and suppose that all the qubits suffer from the same noise. 
Such a noise is called fully correlated and it may happen when 
the dimensions of the quantum computer are microscopic compared with the
wavelength of external disturbances. Noiseless subsystem (NS)
\cite{ns1,ns2,ns3,Kempe} and decoherence free subsystem (DFS)
\cite{dfs1,dfs2,dfs3,dfs4} are well known strategies to protect
a system from such
fully correlated noises \cite{gc,chl}. These schemes,
however, require 
ancillae in pure states and thus they are {\it expensive}. 

In the following, we show that it is indeed possible to
devise a cheaper QEC scheme employing ancillae in the uniformly 
mixed state. 
Let
\begin{eqnarray}
\rho_1 &=& \sigma_0/2 
        + (n_x, n_y, n_z) \centerdot (\sigma_x, \sigma_y, \sigma_z)/2
\label{eq:rho1}
\end{eqnarray}
be the state of the qubit to be protected.
Here $\sigma_0$ is a unit matrix of dimension 2, 
${\bm n} = (n_x, n_y, n_z)$ is the Bloch vector, and $\sigma_i$ is the
$i$th component of the Pauli matrices. 
We introduce two ancillae in uniformly mixed states, whose  
Bloch vectors are
$\bm{0}$. 
The initial state of the three-qubit system is thus a tensor product state
$\rho_1 \otimes (\sigma_0/2)^{\otimes 2}$.
The unitary encoding operator $U_{\rm E}$ transforms the tensor 
product state 
to an entangled state $\tilde{\rho}_3$.
If the state of the system is again $\tilde{\rho}_3$ even after the 
action of
noises, a unitary recovery operator, $U_{\rm R}=U_{\rm E}^\dagger$, 
transforms $\tilde{\rho}_3$ back to the initial tensor product state 
$\rho_1 \otimes (\sigma_0/2)^{\otimes 2}$ and $\rho_1$ can be recovered
after tracing over the ancilla states.

It is highly counterintuitive that a QEC scheme
works with ancillae in uniformly mixed states. 
The trick is that the uniformly mixed state
$ (\sigma_0/2)^{\otimes 2}$ 
is rewritten as  
\begin{eqnarray}
&& \frac{1}{4}\left( |  \bm{n}_2, \bm{n}_2'  \rangle \langle  \bm{n}_2, \bm{n}_2' |
+ |  -\bm{n}_2, -\bm{n}_2'  \rangle \langle  -\bm{n}_2, -\bm{n}_2' | 
\right. \nonumber\\
&+& |  -\bm{n}_2, \bm{n}_2'  \rangle \langle  -\bm{n}_2, \bm{n}_2' | 
 + |  \bm{n}_2, -\bm{n}_2'  \rangle \langle  \bm{n}_2, -\bm{n}_2'|
\left. \right),
\end{eqnarray}
where ${\hm n}_2$ and ${\hm n}_2'$ are arbitrary Bloch vectors 
($|{\hm n}_2| = |{\hm n}_2'| =1$) and 
$|{\bm n}_2 \rangle$ and $|{\bm n}_2' \rangle$ are
pure states corresponding to ${\bm n}_2$ and ${\bm n}_2'$,
respectively. 
If a QEC scheme works with arbitrary pure ancilla states, 
the superposition principle of quantum mechanics 
guarantees that ancillae in a uniformly mixed state do work as well.

A more formal description is given as follows.
Suppose we have a single qubit in a state $\rho_1$, which we
want to protect from noise operators $\{\sigma_x, \sigma_y, \sigma_z\}$.
To this end, we introduce two additional qubits, which may be in an
arbitrary state $\rho_2$, and 
apply a suitable encoding operator $U_E$ on $\rho_1 \otimes \rho_2$ to 
obtain a codeword $\tilde{\rho}_3 = U_E (\rho_1 \otimes 
\rho_2) U_E^{\dagger}$.
We introduce the fully correlated error channel $\Phi$ represented by 
\begin{eqnarray}
\Phi (\tilde{\rho}_3) 
&=& \sum_{i=0}^3 p_i X_i \tilde{\rho}_3 X_i^\dagger,
\label{eq:errorch}
\end{eqnarray}
where $X_0 = \sigma_0^{\otimes 3}, X_1 = \sigma_x^{\otimes 3}, 
X_2 = \sigma_y^{\otimes 3}, X_3 = \sigma_z^{\otimes 3}$.
Here $p_i \ge 0 $ is the probability with which an error operator $X_i$
acts on $\tilde{\rho}_3$ and we assume $\sum_{i=0}^3 p_i =1$.
Suppose there is an encoding operator $U_E$ satisfying
\be
U_E^{\dagger} X_i U_E = \sigma_0 \otimes M_i
\ee
for $i=1, 2, 3$. 
Then, $U_E$ defines the QEC scheme that we are seeking.
We can show that
\begin{eqnarray}\label{eq:qec12}
U_R \Phi
(U_E(\rho_1 \otimes \rho_2) U_E^{\dagger})U_R^{\dagger}
&=& \sum_{i=0}^3 p_i (\rho_1 \otimes  M_i \rho_2 M_i^{\dagger})
\nonumber \\
&=& \rho_1 \otimes \rho_2',\label{eq:aaa}
\end{eqnarray}
where $\rho_2'= \sum_i p_i M_i \rho_2 M_i^{\dagger}$.
This proves that, after decoding,
the error channel $\Phi$ affects only $\rho_2$ but not $\rho_1$.

\begin{figure}[t]
\begin{center}
\includegraphics[width=6cm]{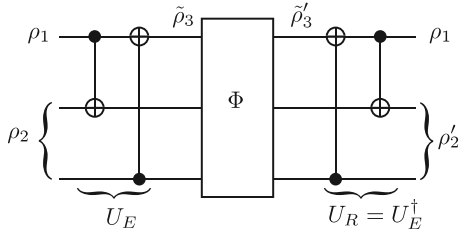}
\end{center}
\caption{
\label{QCircuit}
Encoding circuit $U_E$, error channel $\Phi$ and recovery
circuit $U_R$ in the simplest case. }
\end{figure}

There are infinitely many choices of $U_E$ but careful inspection
of the error operators reveals that 
$U_E= U_{\rm CNOT31}~ U_{\rm CNOT12}$
is the simplest choice \cite{pure_a,scq-QEC}.
The $4 \times 4$ matrices $\{M_i\}$ are obtained by direct
calculation as $M_0= \sigma_0^{\otimes 2}, M_1=\sigma_x^{\otimes 2},
M_2= - \sigma_y \otimes \sigma_x$
and $M_3= \sigma_z \otimes \sigma_0$.
Figure \ref{QCircuit} shows the encoding circuit $U_E$,
the error channel $\Phi$ and the recovery
circuit $U_R$.

\begin{figure}[b]
\begin{center}
\includegraphics[width=7.5cm]{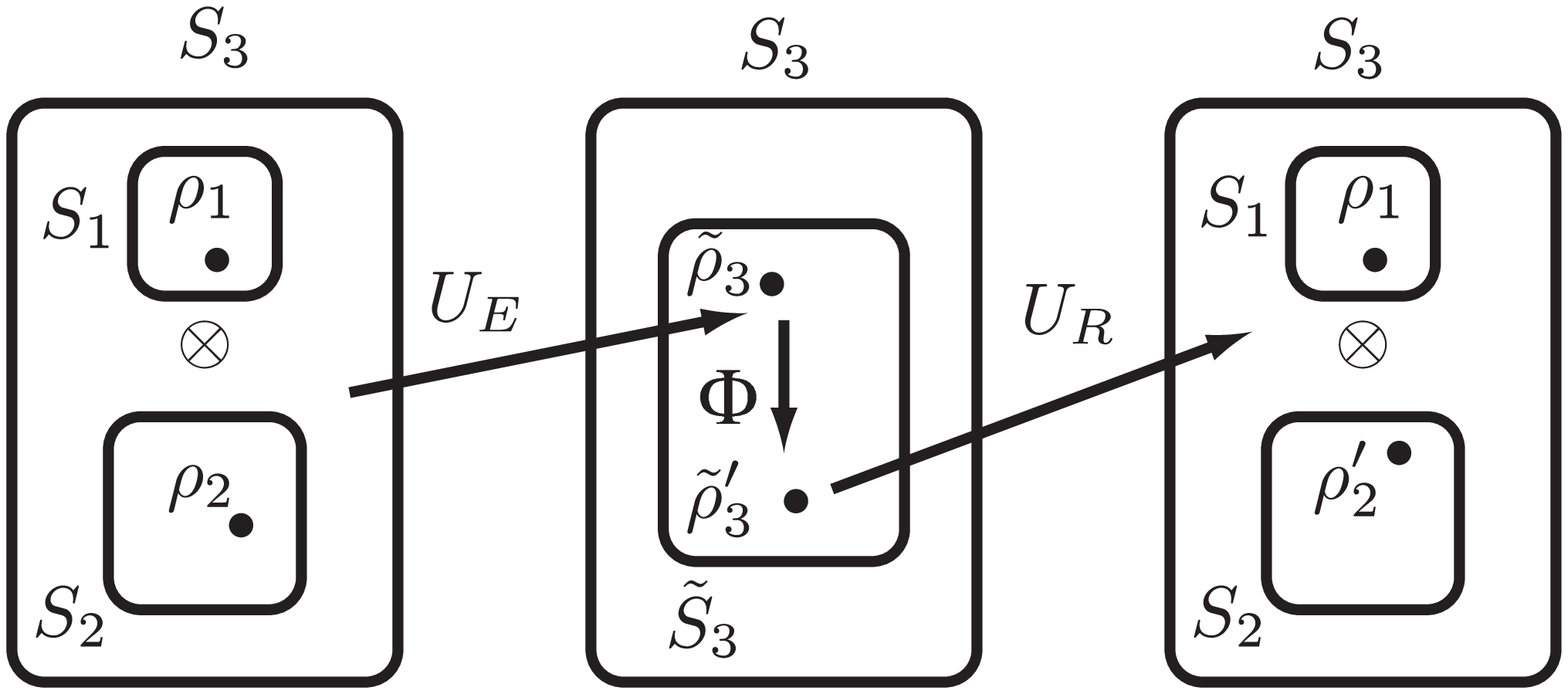}
\end{center}
\caption{
\label{fig2}
Encoding operator $U_E$ maps a tensor product
state $\rho_1 \otimes \rho_2 \in {\mathcal{S}_1} \otimes 
{\mathcal{S}}_2$ to a codeword $\tilde{\rho}_3
\in \tilde{{\mathcal{S}}}_3$.
The error channel ${\Phi}$ maps $\tilde{\rho}_3$ to $\tilde{\rho}_3'$
within the code space $\tilde{{\mathcal{S}}}_3$. As a result, the recovery 
operator
$U_R$ maps $\tilde{\rho}_3'$ to $\rho_1 \otimes \rho_2'$, restoring
the data qubit.
}
\end{figure}

Let $\mathcal{S}_1$ and $\mathcal{S}_2$ be the state spaces 
of the data qubit and the
ancillae, respectively. The set of the encoded states
$\tilde{\mathcal{S}}_3 = U_E (\mathcal{S}_1 \otimes \mathcal{S}_2)
 U_E^{\dagger}$ is a subset
of the total state space $\mathcal{S}_3$ of the three-qubit system.
How our QEC scheme works is summarized in Fig.~\ref{fig2}. 

The extreme case of the uniformly mixed state 
$\rho_2 = \sigma_0^{\otimes 2}/4$
is worth analyzing separately.
When the data qubit is in a pure state, this provides an interesting
example of DQC-1 \cite{dqc1}. Moreover, our QEC scheme 
equally works for a data qubit in a mixed initial state.
It is readily found that 
\be
U_R {\Phi}\left(U_E\left(\rho_1 \otimes \frac{\sigma_0^{\otimes 2}}{4}\right)
U_E^{\dagger}\right)U_R^{\dagger}
= \rho_1 \otimes  \frac{\sigma_0^{\otimes 2}}{4}.
\ee
This shows that $\rho_1 \otimes \sigma_0^{\otimes 2}/4$ is 
a fixed point of this operation
for any $\rho_1 \in \mathcal{S}_1$. 

Let us compare our QEC scheme with the NS encoding scheme discussed in \cite{qecc3}. 
This NS encoding scheme
employs three qubits to encode a logical qubit robust
against any noise of the form $W^{\otimes 3}$,
where $W$ is an arbitrary element of the 2-dimensional
representation of SU(2). 
It was shown for this scheme that
\begin{eqnarray*}
\lefteqn{U'_R 
{\Phi'}\left(U'_E(\rho_1 \otimes |0 \ket\bra 0|\otimes \rho_a
	){U'_E}^{\dagger}
\right) 
{U'_R}^\dagger} \nonumber \\
 &=& \rho_1 \otimes |0 \ket\bra 0| \otimes 
\sum_{i=0}^3 p_i U_i \rho_a U_i^{\dagger},
\end{eqnarray*}
where $U'_E$ and $U'_R$ are the encoding and the recovery operators,
respectively. 
$\{U_i\}=\{ \sigma_0, e^{i \alpha \sigma_x}, e^{i \beta \sigma_y}, 
e^{i \gamma \sigma_z} \}$ is the set of error operators and
$\Phi'$ is defined by an expression analogous to Eq.~(\ref{eq:errorch}). 
For this scheme, the initial ancilla state is required to be of the
form $|0 \ket \bra 0| \otimes \rho_a$. 
In contrast, although the error operators avoidable with our scheme are
restricted within a subset of those avoidable in \cite{qecc3},
our proposal has the remarkable advantage that any initial ancilla
state $\rho_2$ can be employed for successful QEC.

We will discuss quantum discord (hereafter, abbreviated as QD) 
introduced in \cite{QD} in order to analyze another aspect of 
our scheme. 
QD is a measure of non-classical correlations 
between two subsystems of a quantum system. 
Surprisingly enough, it was found that QD may be non-vanishing even in
the absence of entanglement and that in fact, there are useful quantum
algorithms that work with little or no entanglement called
\mbox{DQC-1 \cite{dqc1,Zeeya,DQC1QD}}. 
In other words, separability alone does not imply 
the absence of a quantum nature of the state. 

The left and the right QDs of our encoded state 
\begin{eqnarray}
\tilde{\rho}_3 
&=& \frac{1}{8}\left(
    \sigma_0      \otimes \sigma_0 \otimes \sigma_0 + 
n_x \sigma_x \otimes \sigma_x \otimes \sigma_0         \right. 
\nonumber \\
&+& \left.  
n_y \sigma_y \otimes \sigma_x \otimes \sigma_z      +
n_z \sigma_z \otimes \sigma_0 \otimes \sigma_z    
\right),
\label{eq:e_state}
\end{eqnarray}
are defined respectively as \cite{QD,DVB}
\begin{eqnarray*}
\mathcal{D}(2:1) 
&=& S(\tilde{\rho}_1)-S(\tilde{\rho}_3) +
\tilde{S}(2|1), 
\label{eq:def1}\\
\mathcal{D}(1:2) 
&=& S(\tilde{\rho}_2)-S(\tilde{\rho}_3) +
\tilde{S}(1|2).
\label{eq:def2}
\end{eqnarray*}
Note that they are not necessarily equal to each other \cite{QD}.  
Here,  $S(\rho) \equiv - {\rm Tr}( \rho \log_2 \rho)$ 
is the von Neumann entropy of a density matrix $\rho$. 
The density matrices $\tilde{\rho}_{1}$ 
and $\tilde{\rho}_{2}$ are obtained by tracing over the ancillae and 
data-qubit states, respectively.
We define a projective measurement by a complete set 
of two orthonormal vectors $\displaystyle \{ |\pm \bm{m} \rangle \}$,
which define
$ \displaystyle 
| \pm \bm{m} \rangle \langle \pm \bm{m}| = \frac{\sigma_0 \pm \bm{m} 
\cdot \bm{\sigma}}{2}
$
with 
$|\bm{m}|=1$. 
Let us define the conditional entropy by
$\displaystyle
S_{ \{ |\pm \bm{m} \rangle \} }(2|1) =  \sum_\pm p_\pm S(\rho_{2|\pm})$, 
where 
$\rho_{2|\pm} 
= \langle \pm \bm{m} | \tilde{\rho}_3 | \pm \bm{m} \rangle /p_\pm$
and 
$p_\pm =  {\rm Tr} \langle \pm \bm{m} | \tilde{\rho}_3 |\pm  \bm{m}
\rangle$. 
Then, $\tilde{S}(2|1)$ is defined as 
$\displaystyle \min_{|\bm{m}|=1} S_{\{ | \pm \bm{m} \rangle \}}(2|1)$.
$\tilde{S}(1|2)$ is defined similarly. 

The explicit form of the ancilla state after the measurement of 
the data qubit with a basis $\{|\pm \bm{m} \rangle \}$ is 
\begin{eqnarray*}
\rho_{2|\pm} 
&=& \frac{1}{4}\left(
                 \sigma_0 \otimes \sigma_0 
 \pm  n_x m_x \sigma_x \otimes \sigma_0   \right. \\
&\pm& n_y m_y \sigma_x \otimes \sigma_z 
 \pm  n_z m_z \sigma_0 \otimes \sigma_z    
\left. \right).
\end{eqnarray*}
The corresponding conditional entropy is
\begin{eqnarray*}
S_{\{ |\pm \bm{m} \rangle \}} (2|1) 
&=& 2 - \frac{1}{8}\sum_{j=1}^8 (1+\bm{n}\cdot\bm{n}_j) 
\log_2 (1+\bm{n}\cdot\bm{n}_j),
\end{eqnarray*}
where $\bm{n}_j = (\pm m_x, \pm m_y, \pm m_z)$ are all eight
combinations of three $\pm$.

\begin{figure}[t]
\includegraphics[bb=0 0 1250 1530,width=4cm]{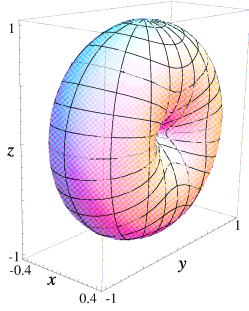}
\caption{Example of $\mathcal{D}_{\{ |\pm \bm{m} \rangle \}}(2:1)$ when 
the initial state of the data qubit has the Bloch vector 
$\bm{n} = \hat{\bm{x}}$. 
$(r, \theta_m, \phi_m)  
= \left(\mathcal{D}_{\{ |\pm \bm{m} \rangle \}}(2:1), \theta_m, \phi_m 
\right)$ is plotted, where $\theta_m$ and $\phi_m$ specify
$\bm{m} = 
(\sin \theta_m \cos \phi_m, \sin \theta_m \sin \phi_m, \cos \theta_m)$. 
$ \mathcal{D}_{\{| \pm \bm{m} \rangle \}}(2:1) =0$ for 
$\bm{m} = \pm \hat{\bm x}$. Therefore, $\mathcal{D}(2:1)$ vanishes for 
$\bm{n} = \hat{\bm{x}}$. 
\label{qd_1}
}
\end{figure}

\begin{figure}[b]
\includegraphics[bb=0 0 550 550,width=5cm]{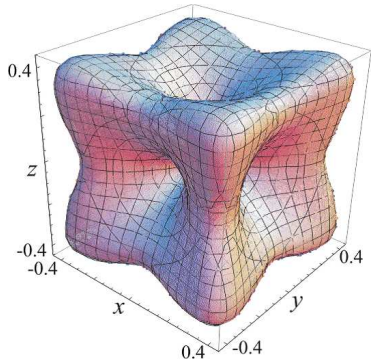}
\caption{Quantum discord $\mathcal{D}(2:1)$ as a function of 
the initial state of the data qubit parameterized by the Bloch vector 
$\bm{n} = (\sin \theta \cos \phi, \sin \theta \sin \phi, \cos \theta)$.
See Eq.~(\ref{eq:rho1}). 
Coordinates $(r=\mathcal{D}(2:1), \theta, \phi)$ depict QD 
as a function of $\theta, \phi$.
$\mathcal{D}(2:1)$ vanishes when 
$\bm{n} = (\pm 1, 0,0), (0,\pm 1, 0), (0,0,\pm 1)$. The function
$\mathcal{D}(2:1)$ takes the maximum value $(3/4)\log_2 3 -1/2$ when 
$\bm{n} = (\pm 1, \pm 1,\pm 1)/\sqrt{3}$.
\label{qd_2}
}
\end{figure}

As an example, we show 
$\mathcal{D}_{\{|\pm \bm{m}\rangle \}}(2:1)
=S(\tilde{\rho_1}) - S(\tilde{\rho_3}) + S_{\{ |\pm \bm{m} \rangle \}} (2|1) $ 
for the initial state $\rho_1$ with the Bloch vector $\bm{n} = \hat{\bm{x}}$
as a function of $\bm{m} = (\sin \theta_m \cos
\phi_m, \sin \theta_m \sin \phi_m, \cos \theta_m)$ in Fig.~\ref{qd_1}.
Here, $\hat{\bm{x}}, \hat{\bm{y}}$ and $\hat{\bm{z}} $ are the 
unit vectors along the $x$-, $y$- and $z$-axes, respectively. 
Note that when $\bm{m} = \hat{\bm x}$,  
$\mathcal{D}_{\{|\pm \bm{m} \rangle \}} =0$. Therefore,
$\mathcal{D}(2:1)=0$.
The quantum discord 
$\mathcal{D}(2:1)$ as a function of $\bm{n}= (\sin \theta \cos
\phi, \sin \theta \sin \phi, \cos \theta)$ is shown in Fig.~\ref{qd_2}.

Although extensive optimization is necessary
to evaluate QD in general, 
some initial states satisfying
$\mathcal{D}(2:1)=0$ are easily obtained 
by carefully inspecting the structure of $\tilde{\rho}_3$. 
Let us consider the case $\bm{n} = \hat{\bm{x}}$, for example.
In this case, $\tilde{\rho}_3$ is reduced to 
$\frac{1}{8}\left(\sigma_0 \otimes \sigma_0 \otimes \sigma_0 + 
\sigma_x \otimes \sigma_x \otimes \sigma_0         \right) $, 
which contains only $\sigma_0$ and $\sigma_x$ for the data qubit. 
Therefore, it is 
reasonable to employ $| \hat{\bm{x}} \rangle $ as a candidate
for $|\bm{m}\rangle$ for obtaining $\tilde{S}(2|1)$. 
With this choice, $\tilde{\rho}_3$ is reduced
to a block-diagonal form  
\begin{eqnarray*}
\tilde{\rho}_3(\bm{n} = \hat{\bm x})
&=& \sum_\pm |\pm \hat{\bm x}\rangle \langle \pm \hat{\bm{x}}| \otimes 
\left( p_\pm \,  \rho_{2|\pm}\right)
\end{eqnarray*}
and
$\mathcal{D}(2:1) = 0$ is readily obtained \cite{QD,DVB}. 

$\mathcal{D}(1:2) = 0 $ for an arbitrary initial state can be 
proved similarly. $\tilde{\rho}_3$ contains only 
$\sigma_0 \otimes \sigma_0$,
$\sigma_x \otimes \sigma_0$, $\sigma_0 \otimes \sigma_z$, and
$\sigma_x \otimes \sigma_z$ for the ancilla qubits. Therefore, we take 
\begin{eqnarray*}
|\Pi_{\pm\pm}\rangle \langle \Pi_{\pm\pm}| &=& 
|\pm \hat{\bm x}, \pm \hat{\bm z} \rangle  \langle \pm \hat{\bm x}, \pm
 \hat{\bm z}|
= \frac{\sigma_0 \pm \sigma_x}{2} \otimes  \frac{\sigma_0 \pm \sigma_z}{2}
\end{eqnarray*}
as a complete set of four unit vectors that determine the projective
measurement on the ancillae, although there are many other
possibilities. 
$\tilde{\rho}_3$ is rewritten as 
\begin{eqnarray*}
\tilde{\rho}_3 &=& \sum_{\pm\pm} \left( p_{\pm\pm} \, \rho_{1|\pm\pm}\right)
\otimes 
|\Pi_{\pm\pm}\rangle \langle \Pi_{\pm\pm}|.
\end{eqnarray*}
When the data qubit and ancillae are rearranged, the density matrix is 
rewritten as a block-diagonal form 
and thus $\mathcal{D}(1:2)=0$ is immediately obtained.

According to the classification introduced in \cite{oppen,horo},
vanishing $\mathcal{D}(1:2)$ implies that our encoded state
has a quantum-classical correlation. Furthermore, in case
$\mathcal{D}(2:1)$ also vanishes, $\tilde{\rho}_3$
has a product eigenbasis as we have shown above, and
the encoded state has a classical-classical correlation, or, in other words,
is (properly) classically correlated.

When the ancillae are pure, we find
$\mathcal{D}(2:1) = \mathcal{D}(1:2)$, which is nothing but the entanglement 
entropy. 
For example, 
\begin{eqnarray*}
&& \mathcal{D}(2:1) = \mathcal{D}(1:2) \\
&=& 2-(1-n_z)\log_2 (1-n_z) -(1+n_z)\log_2 (1+n_z),
\end{eqnarray*}
when 
$\rho_2 = 
|\bm{z}\rangle \langle \bm{z}| \otimes |\bm{z}\rangle \langle \bm{z}| $.

We demonstrate our QEC scheme with a NMR quantum computer. 
We employ a JEOL ECA-500 NMR spectrometer \cite{jeol}, 
whose hydrogen Larmor frequency is approximately 500~MHz. 
We employ a linearly aligned three-spin molecule,
 ${}^{13}$C-labeled L-alanine (98\% purity, Cambridge Isotope) 
solved in D$_2$O.

We simplify the quantum circuit shown in Fig.~\ref{QCircuit} 
by taking into account the fact that the phases of states are not 
independently observed in a NMR quantum computer.  
Both the encoding and the decoding require
only 5 pulses including refocusing pulses, taking approximately 25~ms.

The density matrix of the thermal state is well approximated by
\begin{eqnarray*}
\rho 
&=& (\sigma_0/2)^{\otimes 3}\\
&+&
\frac{\epsilon}{8} \left(\sigma_z \otimes \sigma_0 \otimes \sigma_0 +
\sigma_0 \otimes \sigma_z \otimes \sigma_0  
+  \sigma_0 \otimes \sigma_0 \otimes \sigma_z\right),
\end{eqnarray*}
where $\epsilon \sim 10^{-6}$.
Since $(\sigma_0/2)^{\otimes 3}$ is not visible in NMR, 
the density matrix of the thermal state 
is considered as a pseudo-pure
state for DQC-1.

\begin{figure}[t]
\begin{center}
\vspace{3ex}
\includegraphics[bb=300 200 2300 970,width=8.5cm]{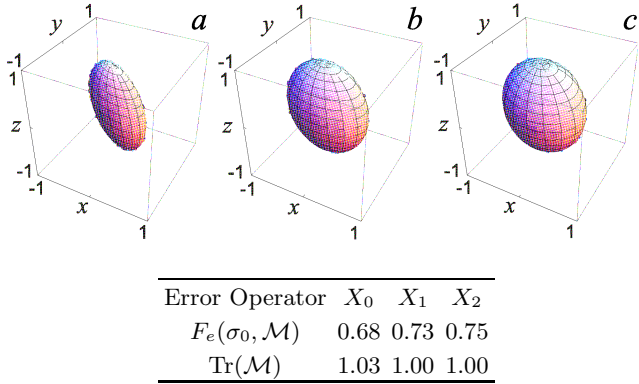}

\vspace{3ex}
\begin{tabular}{cccc}
\hline
Error Operator & $X_0$& $X_1$ & $X_2$ \\
$F_e(\sigma_0,{\mathcal M})$ & $0.68$& $0.73$ & $0.75$\\
${\rm Tr}({\mathcal M})$ & $1.03$& $1.00$ & $1.00$\\
\hline
\end{tabular}
\end{center}
\caption{
\label{fig:tomography}
Visualization of error correction performances. The surface of the 
Bloch sphere is mapped onto the surfaces in (a), (b) and (c) corresponding
 to three different error operators, $X_0, X_1, X_2$, respectively. 
See the text for details. Entanglement fidelities
 $F_e(\sigma_0,\mathcal{M})$ and traces ${\rm Tr}({\mathcal M})$ are summarized in 
the table. ${\mathcal M}$ 
represents a map which is determined by
the encoding, error, and recovery processes depicted in Fig.~\ref{QCircuit}. 
}
\end{figure}

We perform three sets of experiments, in which we set
\begin{eqnarray*}
\begin{array}{ccc}
(a) & \{p_i\} = (1,0,0,0) & : {\rm no~error} \\
(b) & \{p_i\} = (0,1,0,0) & : X_1{\rm ~error} \\
(c) & \{p_i\} = (0,0,1,0) & : X_2{\rm ~error}, 
\end{array}
\end{eqnarray*}
respectively, in Eq.~(\ref{eq:errorch}). We do not need to examine 
the $X_3$ error
separately since $X_3 = i X_2 ~ X_1$.
Each set starts with 4 different initial states in order to 
apply quantum process tomography \cite{kondo}. 
The results are summarized in Fig.~\ref{fig:tomography}.
Although the surfaces are
distorted, it is clear that our QEC scheme indeed eliminates
the effects of the fully correlated noises. 
The table in Fig.~\ref{fig:tomography} summarizes 
the entanglement fidelities.

It is noteworthy that one-qubit gate operations on
the logical qubit take simple forms. Let $V$ be a one-qubit
gate acting on the logical qubit. Then its action on 
the physical qubits is obtained by simplifying 
$U_E (V \otimes \sigma_0 \otimes \sigma_0) U_R$.
For the simple gates $V=\sigma_x, \sigma_y$ and $\sigma_z$,
the corresponding operations on the physical qubits are
$ \sigma_x \otimes \sigma_x \otimes \sigma_0, 
\sigma_y \otimes \sigma_x \otimes \sigma_z, $
and $\sigma_z \otimes \sigma_0 \otimes \sigma_z, 
$ respectively. Note that these operators 
satisfy the ordinary $\mathfrak{su}(2)$ algebra.
It is easy to obtain more general gate operations acting on the physical
qubits by simply exponentiating these operators,
e.g., $e^{-i \alpha \sigma_x \otimes \sigma_x \otimes \sigma_0}$ 
implements $V = e^{-i \alpha \sigma_x}$.
Note that $e^{-i \beta \sigma_y \otimes \sigma_x \otimes \sigma_z}=
e^{-i \pi (\sigma_z \otimes \sigma_0 \otimes \sigma_z)/4}
e^{-i \beta \sigma_x \otimes \sigma_x \otimes \sigma_0} 
e^{i \pi (\sigma_z \otimes \sigma_0 \otimes \sigma_z)/4}$
in the case of $V=e^{-i \beta \sigma_y}$. From these operators,
we can understand how the information of the data qubit 
is distributed in the encoded state. 
We note that direct operations on logical qubits in DFS/NS were
discussed in \cite{gates_in_DFS}. 

In summary,
we demonstrated a quantum error correction scheme 
avoiding fully correlated errors, 
in which the ancillae can be
in a uniformly mixed state. 
Our results pave the way to new applications of DQC-1 to quantum
computing. The analysis of quantum discord reveals that 
our encoding creates an interesting quantum correlation between 
the data qubit and the 
ancilla qubits; our encoded state has a quantum-classical
correlation in general and has a classical-classical correlation 
when both left and right quantum discords vanish. 
We anticipate further progress both in the
understanding of quantum correlations and the
development of QEC schemes. Our QEC scheme admits simple
one-qubit gate operations on the encoded qubits.

We are grateful to Hiroyuki Tomita for his valuable
inputs and to Akira SaiToh for his critical reading. 
We are also grateful to the `Open Research Center' Project for 
Private Universities,
matching fund subsidy from the MEXT (Ministry of Education, 
Culture, Sports, Science
and Technology) for financial support. Y.\ K.\ and M.\ N.\ would like to 
thank partial
supports of Grants-in-Aid for Scientific Research from the
JSPS (Grant No. 23540470).
C.\ B.\ is supported by the MEXT Scholarship for
foreign students.


\begin{thebibliography}{9}

\bibitem{gaitan}
F.\ \ Gaitan, {\it Quantum Error Correction and Fault Tolerant Quantum Computing}
(CRC Press, New York, 2008).

\bibitem{divin}
D. P.\  DiVincenzo, 
Fortschritte\ der\ Physik\ {\bf 48}, 771 (2000).

\bibitem{pure_a}
B.\  Criger, O.\ Moussa and R.\ Laflamme,
Phys.\ Rev.\ A {\bf 85}, 044302 (2012).

\bibitem{qecc3} 
C.-K.\ Li, M.\ Nakahara, Y.-T.\ Poon, N.-S.\ Sze and H.\
Tomita,
Phys.\ Rev.\ A {\bf 84}, 044301 (2011).

\bibitem{QD}
H.\ Ollivier and W. H. Zurek, Phys.\ Rev.\ Lett.\ {\bf 88}, 017901 (2001),
L.\ Henderson and V.\ Vedral, J.\  Phys. A {\bf 34},  6899 (2001).

\bibitem{dqc1} 
E.\ Knill and R.\ Laflamme, Phys.\ Rev.\ Lett.\ {\bf 81},
5672 (1998).

\bibitem{Zeeya}
Z.\ Merali, Nature, {\bf 474}, 24 (2011). 

\bibitem{ns1} 
E.\ Knill, R.\ Laflamme and L.\ Viola,
Phys.\ Rev.\ Lett., {\bf 84}, 2525 (2000).

\bibitem{ns2} 
S.\ De Filippo, Phys.\ Rev.\ A {\bf 62}, 052307 (2000).

\bibitem{ns3} 
C.-P.\ Yang and J.\ Gea-Banacloche, Phys.\ Rev.\ A {\bf 63},
022311 (2001).

\bibitem{Kempe} 
J.\ Kempe, D.\ Bacon, D.\ A.\ Lidar and K.\ B.\ Whaley,
Phys.\ Rev.\ A {\bf 63}, 042307 (2001). 

\bibitem{dfs1} 
P.\ Zanardi and M.\ Rasetti, Phys.\ Rev.\ Lett., {\bf 79}, 3306
(1997).

\bibitem{dfs2} 
P.\ Zanardi and M.\ Rasetti, Mod.\ Phys.\ Lett.\ B
{\bf 11}, 1085 (1997).

\bibitem{dfs3} 
P.\ Zanardi, Phys.\ Rev.\ A {\bf 57}, 3276 (1998).

\bibitem{dfs4} 
D.\ A.\ Lidar, I.\ L.\ Chuang and K.\ B.\ Whaley,
Phys.\ Rev.\ Lett., {\bf 81}, 2594 (1998).

\bibitem{gc}
G.\ Chiribella, M.\ Dall'Arno, G.M.\ D'Ariano, C.\ Macchiavello, 
P.\ Perinotti, Phys.\ Rev.\ A {\bf 83}, 052305  (2011).

\bibitem{chl}
C.-K.\ Li, M.\ Nakahara, Y.-T.\ Poon, N.-S.\ Sze, H.\ Tomita
Phys.\ Lett.\ A {\bf 375}, 3255 (2011). 

\bibitem{scq-QEC}
M.\ D.\ Reed, L.\ DiCarlo, S.\ E.\ Nigg, L.\ Sun, L.\ Frunzio,
S.\ M.\ Girvin and R.\ J.\ Schoelkopf,
Nature {\bf 482}, 382 (2012). 

\bibitem{DQC1QD}
A.\ Datta, A.\ Shaji, and C.\ M.\ Caves, Phy.\ Rev.\ Lett.\ {\bf 100}, 
 050502 (2008). 

\bibitem{DVB}
B.\ Daki\'c, V.\ Vedral and C.\ Brukner,
Phys.\ Rev.\ Lett.\ {\bf 105}, 190502 (2010). 

\bibitem{oppen} 
J.\ Oppenheim, M.\ Horodecki, P.\ Horodecki and R.\ Horodecki, Phys.\ Rev.\ Lett. 
{\bf 89}, 180402 (2002).

\bibitem{horo} M.\ Horodecki, P.\ Horodecki, R.\ Horodecki, J.\ Oppenheim, 
A.\ Sen(De), U.\ Sen and B.\ Synak-Radtke, Phys.\ Rev.\ A {\bf 71}, 062307 (2005).

\bibitem{jeol}
http://www.jeol.com/. 

\bibitem{kondo} 
See, for example, Y.\ Kondo, 
{J.\ Phys.\ Soc.\ Jpn.} {\bf 76},  104004 (2007) and references therein. 

\bibitem{ee} H.\ Barnum, M.\ A.\ Nielsen, and B.\ Schumacher, Phys.\ Rev.\ A
{\bf 57}, 4153 (1998).

\bibitem{gates_in_DFS} C.\ A.\ Bishop and M.\ S.\ Byrd, 
J.\ Phys.\ A: Math.\ Theor.\ {\bf 42}, 055301  (2009).


\end{thebibliography}
\end{document}